\documentclass[lettersize,journal]{IEEEtran}
\usepackage{amsmath,amsfonts}
\usepackage{algorithmic}
\usepackage{algorithm}
\usepackage{array}
\usepackage[caption=false,font=normalsize,labelfont=sf,textfont=sf]{subfig}
\usepackage{textcomp}
\usepackage{stfloats}
\usepackage{url}
\usepackage{verbatim}
\usepackage{graphicx}
\usepackage{cite}
\usepackage{url}
\usepackage{multirow}
\usepackage{xcolor}
\usepackage{svg}
\usepackage{threeparttable}
\renewenvironment{IEEEbiography}[1]
  {\IEEEbiographynophoto{#1}}
  {\endIEEEbiographynophoto}

\usepackage{xspace}
\newcommand{\eg}{\textit{e.g.},\xspace}
\newcommand{\ie}{\textit{i.e.},\xspace}

\usepackage{tikz}
\newcommand\publishedtext{%
    \footnotesize This is the authors' accepted version of the article. The final version published by IEEE is J. Hou , S. Bakirtzis, K. Qiu, S. Liao, H. Song, H. Hu, K. Wang, and J. Zhang, `iPLAN: Redefining Indoor Wireless Network
    Planning Through Large Language Models", IEEE Communications Magazine, vol. TBD, no. TBD, pp. TBD, doi:  TBD.} % doi to be fixed
\newcommand\copyrighttext{%
  \footnotesize \textcopyright 2025 IEEE. Personal use of this material is permitted.
  Permission from IEEE must be obtained for all other uses, in any current or future
  media, including reprinting/republishing this material for advertising or promotional
  purposes, creating new collective works, for resale or redistribution to servers or
  lists, or reuse of any copyrighted component of this work in other works.}
\newcommand\copyrightnotice{%
\begin{tikzpicture}[remember picture,overlay]
\node[anchor=north,yshift=0pt] at (current page.north) {\fbox{\parbox{\dimexpr\textwidth-\fboxsep-\fboxrule\relax}{\publishedtext}}};
\node[anchor=south,yshift=10pt] at (current page.south) {\fbox{\parbox{\dimexpr\textwidth-\fboxsep-\fboxrule\relax}{\copyrighttext}}};
\end{tikzpicture}%
}
\begin{document}
\title{iPLAN: Redefining Indoor Wireless Network Planning Through Large Language Models}

\author{Jinbo~Hou,
        Stefanos~Bakirtzis,~\IEEEmembership{Member, IEEE},
        Kehai~Qiu,~\IEEEmembership{Member, IEEE,}
        Sichong~Liao,
        Hui Song,
        Haonan~Hu,
        Kezhi~Wang,~\IEEEmembership{Senior Member, IEEE,}
        Jie~Zhang,~\IEEEmembership{Senior Member, IEEE}
 \thanks{Kezhi Wang and Jie Zhang are corresponding authors.}
 %Jinbo Hou and Jie Zhang are with the Department of Electronic and Electrical Engineering, the University of Sheffield, Sheffield, UK. Stefanos Bakirtzis and Kehai Qiu are with the Department of Computer Science and Technology, University of Cambridge, Cambridge CB3 0FD, U.K. (kq218@cam.ac.uk; ssb45@cam.ac.uk), Kehai Qiu is also with Brunel University London. Kezhi Wang is with the Department of Computer Science, Brunel University London, Uxbridge, Middlesex, UB8 3PH (email: kezhi.wang@brunel.ac.uk).}
}

% The paper headers
%\markboth{IEEE Communications Magazine,~Vol.~xx, No.~xx, June~2025}%
%{Shell \MakeLowercase{\textit{et al.}}: Bare Demo of IEEEtran.cls for IEEE Journals}

\maketitle

\copyrightnotice\vspace*{-8pt} 

\begin{abstract}

Efficient indoor wireless network (IWN) planning is crucial for % enhancing user experience and 
providing high-quality 5G in-building services.
However, traditional meta-heuristic and artificial intelligence-based planning methods face significant challenges due to the intricate interplay between indoor environments (IEs) and IWN demands.
% framework   
In this article, we present an indoor wireless network Planning with large LANguage models (iPLAN) framework, which integrates multi-modal IE representations into large language model (LLM)-powered optimizers to improve IWN planning.
First, we instate the role of LLMs as optimizers, outlining embedding techniques for IEs, and introducing two core applications of iPLAN: ($i$) IWN planning based on pre-existing % pre-existing 
IEs and ($ii$) joint %planning 
design of IWN and IE for new wireless-friendly buildings.
% framework modules
For the former, we embed essential information into LLM optimizers by leveraging indoor descriptions, domain-specific knowledge, and performance-driven perception. 
% 步步推理
For the latter, we conceptualize a multi-agent strategy, where intelligent agents collaboratively address key planning sub-tasks in a step-by-step manner while ensuring optimal trade-offs between the agents.
% simulation
The simulation results demonstrate that iPLAN achieves superior performance in IWN planning tasks and optimizes building wireless performance through the joint design of IEs and IWNs, exemplifying a paradigm shift in IWN planning.

\end{abstract} 


\section{Introduction}

% 下标添加prompt  github

\IEEEPARstart{E}{ffective} indoor wireless network (IWN) planning is fundamental to establishing a robust wireless ecosystem and delivering high-quality indoor communication services in 5G networks \cite{indoor_traffic}. 
However, optimal IWN planning is a highly intricate and multi-dimensional problem that lacks a closed-form or a one-size-fits-all solution due to the complexity of IWNs.
% reasons
%(\eg various physical propagation environments, the existing network topology, hardware specifications, user demands, and the balance between diverse objectives), 
Firstly, IWNs are characterized by the ultra-dense deployment of access points (APs) and the co-existence of different wireless communication technologies, \eg cellular network and WiFi, requiring tailored spectrum and interference management.
Moreover, IWN mobile service demands fluctuate intensely over time and across different indoor environments (IEs), requiring adaptive dynamic resource allocation schemes \cite{bakirtzis2023characterizing}.
Furthermore, indoor radio propagation presents additional difficulties compared to outdoor scenarios, as wireless signals encounter multiple obstacles, leading to pronounced multi-path effects and severe fading.

% meta-heuristic 
Hence, it is imperative to develop optimization frameworks tailored to IEs, particularly in light of the imminent wide-scale IWN deployment in 5G systems \cite{indoor_traffic}. 
Traditionally, IWN planning has relied on meta-heuristic optimization algorithms and artificial intelligence (AI)-related optimization methods, such as reinforcement learning (RL) and deep learning (DL) \cite{Indoor_Optimal_Helmholtz, Planning_DeepRay}. In these cases,  the potential deployment space is explored and refined iteratively based on evaluations derived from site-specific radio propagation models.
However, these approaches face notable limitations: ($i$) they require high-quality training datasets and extensive online fine-tuning, ($ii$) they are prone to being trapped in local optima when addressing multi-objective or constraint-laden IWN planning problems, resulting in inferior performance, and ($iii$) their computational complexity is high, often necessitating a large number of iterations to achieve convergence.

The rapid advances in generative AI and large language models (LLMs) present an unprecedented opportunity to address these limitations and holistically revise IWN planning. Specifically, recent works have demonstrated the effectiveness of LLMs in comprehending complex optimization problems through natural language prompts and established their potential to \textit{act as optimizers}  \cite{LLM_As_Optimizers}. 
Such LLM-powered optimizers possess extensive knowledge in various domains, can adapt fast  % pre-trained various domain knowledge
via few-shot learning, and exhibit impressive reasoning capabilities in optimization problems \cite{Optimiz}. Hence, several works have emphasized the pivotal role of LLMs in various aspects of the wireless ecosystem \cite{LLM_Next_Big_Thing}, multi-agent systems \cite{multi-agent}, wireless intelligence \cite{WirelessLLM}, and multi-modal semantic communications \cite{semantic}. 
However, none of the prior LLM-based frameworks have tackled the intricate challenges arising from the interplay between architecture and wireless in IWN planning. Consequently, this raises a crucial question: \textit{How can LLMs be exploited for IWN planning?}

% overall
Our work answers this question, shedding light on the role of LLMs in the IWN ecosystem and providing the first comprehensive assessment of the techniques that enable LLMs to underpin cutting-edge IWN planning strategies.
Hence, in contrast to previous works, this paper shares our vision of integrating LLM-powered optimizers in next-generation IWNs to support automated network deployment by harnessing multi-modal information related to IEs, \eg building layout and material characteristics.
% works
To illustrate this transformative approach, we detail key features and introduce a general framework for \textbf{i}ndoor wireless network \textbf{P}lanning with large \textbf{LAN}guage models (iPLAN) that can adapt to handle different IWN deployment cases. 
% case
Specifically, we discuss and showcase the effectiveness of iPLAN for two illustrative cases: AP deployment to fulfill indoor coverage for pre-existing IE, and joint IE and IWN design, aiming at inaugurating a new era of wireless-friendly buildings.
% result
The results indicate that the proposed iPLAN framework can significantly improve IWN performance and building's wireless performance.

\begin{figure*}[t]
    \centering
    \includegraphics[width=1.75\columnwidth]{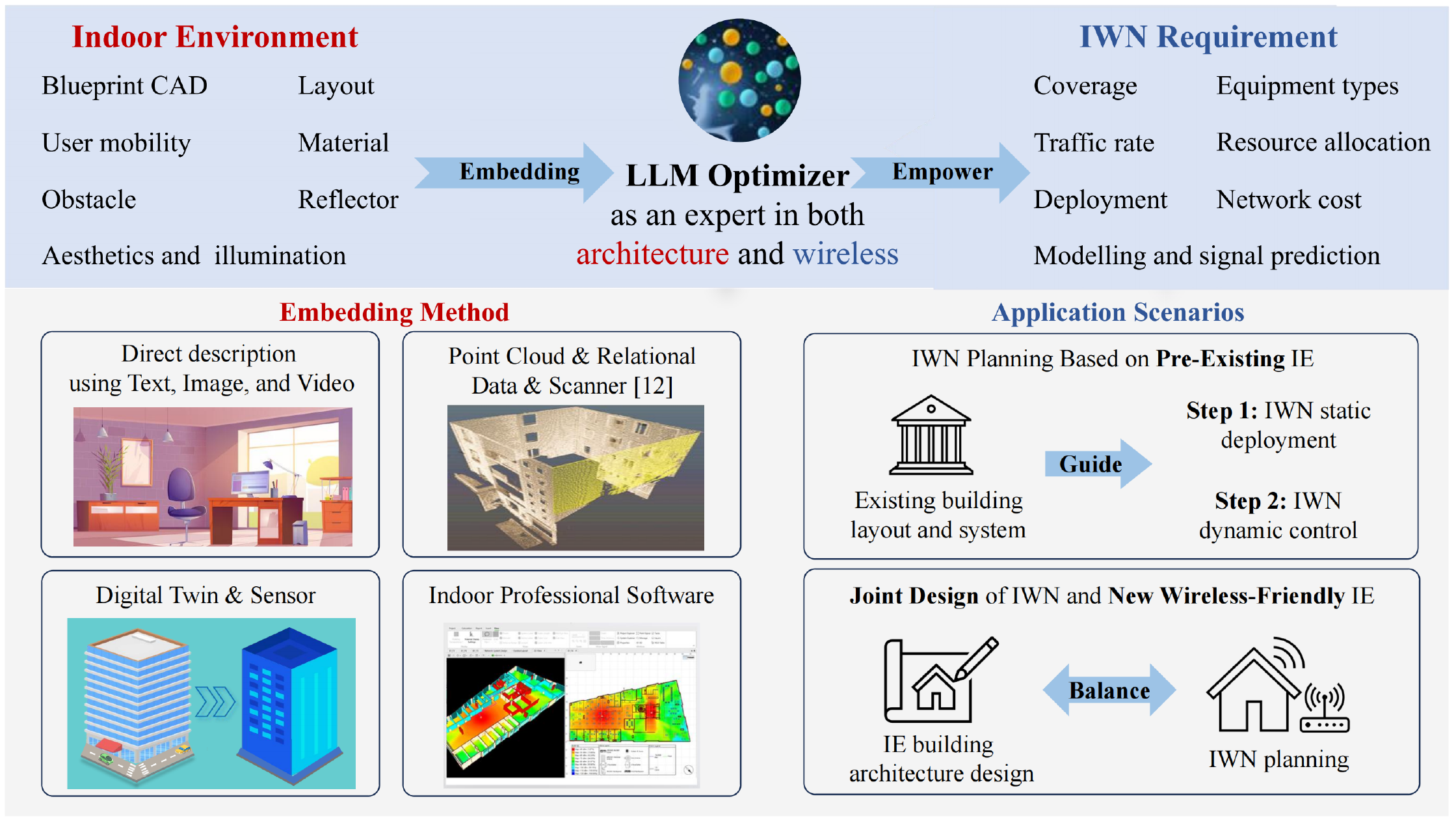}
        \vspace{-1mm}
    \caption{Interplay between LLM optimizers, IEs, and IWNs, including embedding methods for buildings and illustrative IWN planning scenarios. Indoor professional software showcase uses the Ranplan Wireless.}
    \label{framework}
\end{figure*}

\section{The Interplay between LLMs and IWN Planning}

In the following subsections, we demonstrate how LLMs act as optimizers by leveraging the embedded information of IEs to support IWN planning, as depicted in Fig. \ref{framework}.

\subsection{LLMs as Optimizers for IWN Planning}

% nutshell
LLMs are transformer-based natural language processing models comprising a vast number of parameters, which are trained on textual data collected from a large variety of sources based on \textit{self-attention} mechanism \cite{WirelessLLM}. 
%This mechanism enables LLMs to handle long sequences of tokenized information and develop a thorough understanding of input data in natural language.
% LLM as optimizer
Recent works revealed strategies that enable LLMs to act as optimizers through \textit{meta-prompts} and solve \textit{black-box} problems, which are prevalent in IWN optimization problems \cite{LLM_As_Optimizers}. Based on the input background and task information, LLMs can understand, infer, and propose optimization solutions, benefiting from their pre-trained knowledge and extra fine-tuning.
These LLM-powered optimizers are characterized by three key attributes:

% % conventional methods
% Conventional methods (\eg convex optimization, gradient-based techniques, and stochastic algorithms) and AI-related methods (\eg RL and DL) are inadequate to solve such black-box IWN optimization problems in terms of effectiveness, time complexity, and generalization.

% % heuristic
% Moreover, the meta-heuristic algorithm designates a strategy to traverse the optimization space based on iterative observations of the values of $f$ at different points with $N$ variables, $\mathbf{X} = [x_1, x_2, ..., x_N]$, of the optimization space to find a nearly-optimal solution.

% \begin{figure*}[h]
%     \centering
%     \includegraphics[width= 1.5\columnwidth]{Figure/LLM as optimizer.pdf}
%         \vspace{-3mm}
%     \caption{Illustration of meta-prompt-based LLM optimizers.}
%     \label{LLM_as_optimizer}
% \end{figure*}

\textbf{Reasoning \& Reflecting Abilities:}
LLM optimizers exhibit a strong reasoning ability within specialized areas.
In particular, their step-by-step reasoning approach allows decomposing complex problems into smaller manageable sub-tasks to derive a final solution.
Especially in multi-objective planning problems, LLM optimizers can simultaneously accommodate multiple key performance indicator (KPI) requirements \cite{multi-obj_2} and strike an optimal balance between competing objectives.   
% reflection
Furthermore, by learning from past solution-feedback results, LLM optimizers continuously refine their decision-making process, iteratively improving solutions over time.

\textbf{Multi-modal Data Alignment:}
In addition to textual data, LLM optimizers can exclusively comprehend and process various data modalities, including video, image, and audio data through feature extraction and cross-modal alignment.
% indoor
For example, the IE, represented via layout images, blueprint computer-aided design (CAD) files, and even room tour videos, can provide rich planar and spatial architectural information. 
% wireless 
Similarly, IWN performance metrics, such as pathloss heatmaps, coverage performance figures, user trajectories, and equipment KPIs can provide diverse wireless insights.
% 融合与冗余
These multi-modal data sources capture the same events and objects from different perspectives, enabling LLM optimizers to leverage multi-dimensional knowledge for enhanced decision-making in IWN planning.

\textbf{Superior Generalization:} 
LLM optimizers leverage extensive pre-trained knowledge and additional multi-modal information through instruction prompts, references from retrieval augmented generation (RAG), and fine-tuning, ensuring a cohesive and consistent understanding of each distinct planning task. That enables LLM optimizers to adapt rapidly to diverse IWN planning scenarios and make well-informed decisions.
% 泛化性
By iteratively refining solutions and accumulating knowledge, a well-designed meta-prompting approach can be applied to a broad range of optimization problems and has hitherto demonstrated remarkable results \cite{Optimiz}.

% \textbf{Dynamic Adaptability:}
% % 1. in context learning 
% Intrinsically, LLM optimizers are endowed with big data analytics capabilities to identify patterns and trends of IWNs in historical datasets.
% % 2. fine-tuning     
% Moreover, LLM optimizers can make well-informed decisions in real-time situations after online fine-tuning with few shots in a new IE \cite{dynamic_LLM}.
% % % modelling example
% % For example, contrary to conventional model-based radio propagation solvers, \eg ray-tracing, LLM optimizers can peruse historical datasets and correlate them to CAD designs of buildings and user behaviour to rapidly predict signal propagation in dynamic IEs.
% Thus, LLM optimizers enable effective IWN prediction (\eg data-driven modelling) and then dynamic control the IWN resources, requiring minimal training cost and fewer shots.

\subsection{Embedding IE into LLM Optimizers}
% information style    automation  cost

Digitizing IEs is essential to provide LLM optimizers with informative and multi-modal details related to indoor geometries, including different types of environmental data. In addition to direct descriptions through text, layout images, and overview videos, we highlight three other methods for embedding IEs into LLM optimizers:

\textbf{Point Cloud \& Scanner:}
Point clouds \cite{indoor_embed} can describe the contour, boundaries, locations, and other attributes (\eg colour, material, and reflection coefficient) of building layouts and entities (\eg windows, doors, and furniture), using discrete points obtained from laser, acoustic, or visual-based scanners.  
These point data and relational data collectively provide LLM optimizers with sufficient and precise IE information.

\textbf{Digital Twin \& Sensor:}
Digital twins create virtual replicas of physical building environments by integrating real-time sensor data.
The superiority of the digital twin lies in its real-time data processing, seamless integration, and predictive capabilities, providing LLM optimizers with time-flexible IE data.
However, the construction of digital twins requires substantial manual effort and incurs additional costs associated with the deployment of sensors.

\textbf{Indoor Professional Software:} A cost-efficient approach is to develop IE models in dedicated professional simulation software related to IWN planning.
The architecture-domain software (\eg Sketch-up, V-Ray, and Rhino) and IWN-domain software (\eg Ranplan Pro and NetSpot) can endow LLM optimizers with rich IE data and additional professional and mathematical capabilities.
However, offline measurement and manual operations are mandatory to build the models of IEs.

\begin{figure*}[t]
    \centering
    \includegraphics[width= 1.70\columnwidth]{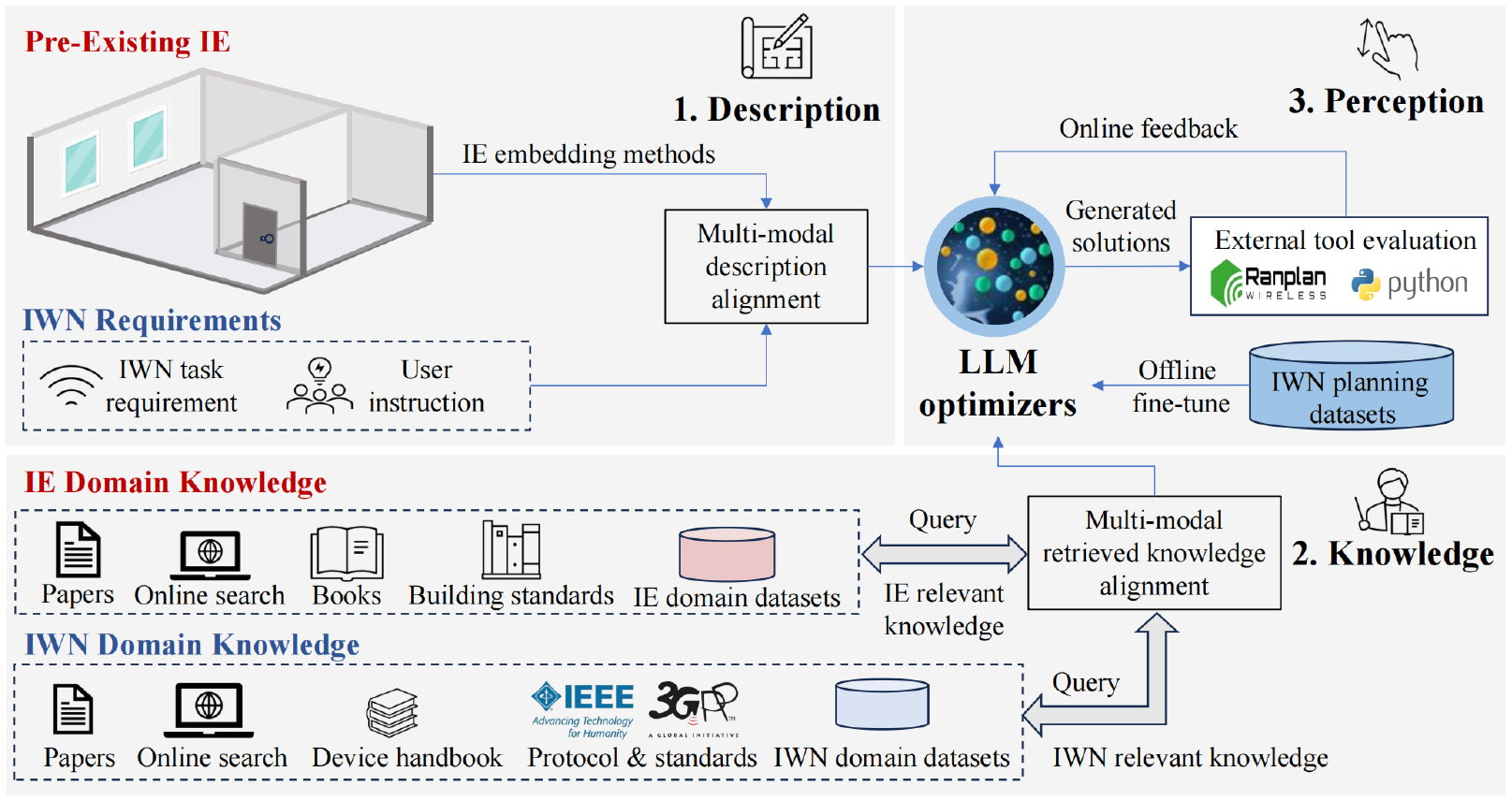}
    \vspace{-2mm}
    \caption{The design of iPLAN framework for IWN planning based on pre-existing IE.}
    \vspace{-2mm}
    \label{IPLAN}
\end{figure*}

\subsection{Applications of LLM Optimizers for IWN Planning}

\textbf{IWN Planning Based on Pre-Existing IE:}
Most IWNs are deployed within pre-existing IEs, \ie buildings that are already constructed without IWN infrastructure. Existing IE-related data can guide LLM optimizers and impose constraints on their decisions. Specifically, the possible deployment options of IWN equipment are significantly affected by the building power supply system, the spatial layout of walls, the existence of obstacles, and other objects within buildings (\eg windows, materials, and doors). 
% % user intend
% Additionally, the UE's intent and requirements should also be considered according to interviews and survey questionnaires.
% fully-constructed indoor buildings
To address these challenges, the multi-objective reasoning capabilities of LLM optimizers can be leveraged to propose a rational deployment solution after embedding diverse multi-modal information from the target IE.
% resource control  动态  重要    
Thereafter, to realize effective dynamic control, LLM optimizers can further analyze traffic trends and infer appropriate resource management strategies via few-shot learning.  
%leveraging their strong adaption abilities through %in-context learning or fine-tuning methods with 

\textbf{Joint Design of IWN and IE for New Wireless-Friendly Buildings:}
%BWP-related
During the building construction phase, the joint IE and IWN design can proactively improve the building wireless performance (BWP) \cite{paper_zhang2022BWP1}. However, the KPIs required by IWNs and IEs may conflict strongly (\eg the equipment deployment and aesthetic effects in IoT services). Hence, LLM optimizers must comprehend and balance the parameters affecting each KPI. 
% 解决问题： 
To address this problem, a promising approach is to utilize multi-agent LLM optimizers to design different parts of IWNs and IEs, leveraging the step-by-step reasoning capacity of LLM optimizers after decomposing into sub-tasks (\eg layout and entity planning, material selection, and IWN planning of various services).
% 合理性评价和evaluation
The generated solutions of LLM optimizers are subsequently evaluated and reported back to the agents for reference and further updates. %until the convergence or maximum steps are attained.

\section{The iPLAN Framework}

This section discusses the iPLAN perspective for pre-existing IEs and describes how it can be adapted to comprise multi-agents for the joint design of IEs and the respective in-building networks. 

\subsection{iPLAN Framework for IWN Planning Based on Pre-Existing IEs} \label{sec:Pre_Existing}
As shown in Fig. \ref{IPLAN}, the iPLAN framework aims at embedding essential IE and IWN information \cite{WirelessLLM} into LLM optimizers and iteratively improving the rationality of IWN planning until certain KPIs are met.
% model  evaluation
In each iteration step, the generated solutions or configuration files are forwarded to an ``evaluator" that models the network performance based on environmental parameters via external simulation tools.
% feedback
Consequently, based on the evaluator's feedback, the LLM optimizer refines its search strategy and the optimization solution until convergence or the maximum number of iterations is reached. This process entails three key modules:

\textbf{Description:}
The description is intended to render LLM optimizers environmental- and problem-aware through multi-modal information, consisting of two types of information. First, IWN contextual information includes building layout, construction materials and their electromagnetic properties, available wireless communication equipment, signal strength between transmitter and receiver, and the physical topology of network architecture.
% 2. task
Second, it comprises details linked to the IWN optimization task, such as the optimization variables, \eg the number of devices, locations, and phase shifter configuration, constraints on shared communication resources (frequency, time, and power), and the optimization target KPIs, such as traffic throughput, coverage, delay, reliability, or monetary cost-efficiency.

\begin{figure*}[t]
    \centering
    \includegraphics[width= 1.85\columnwidth]{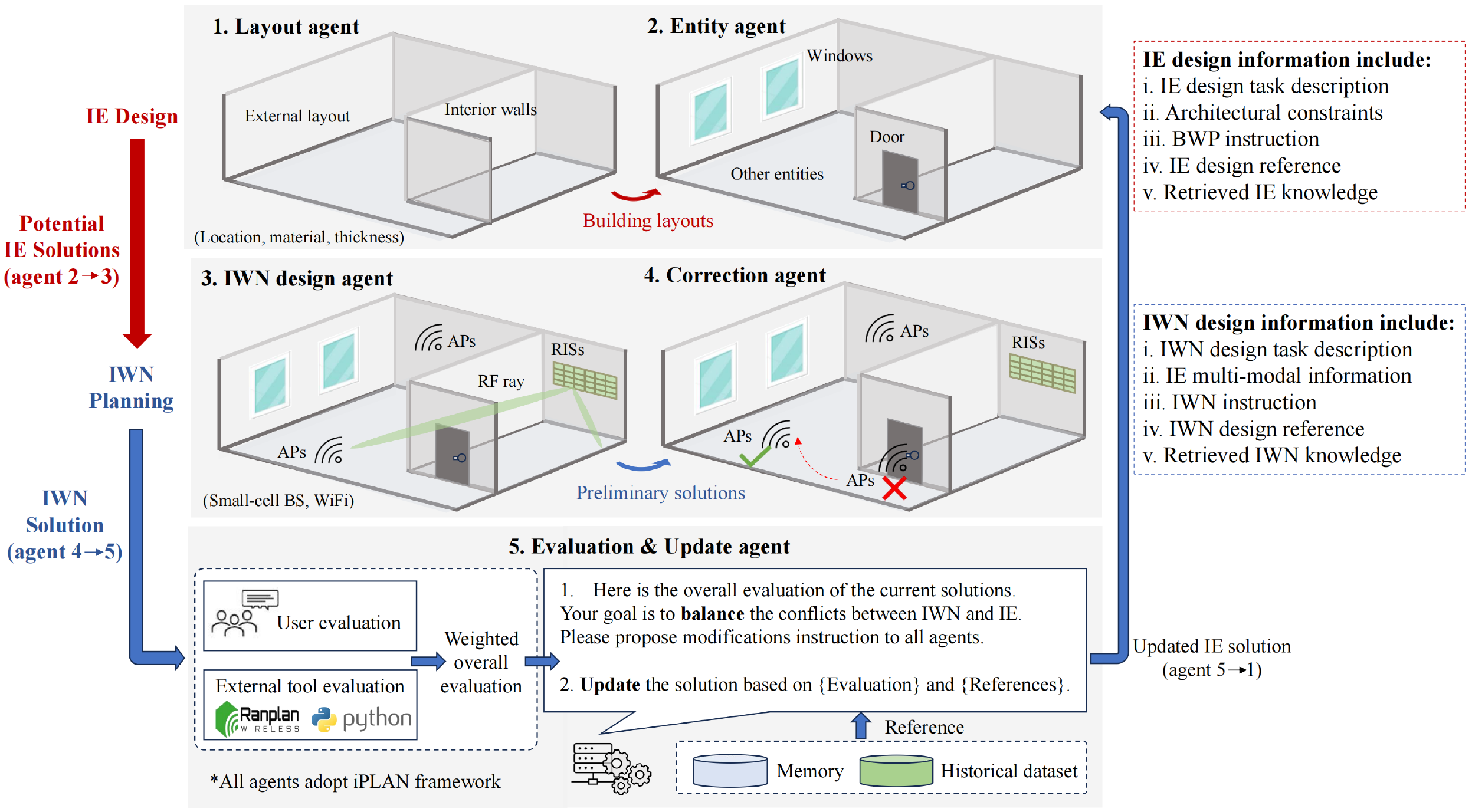}
    \vspace{-1mm}
    \caption{The design of multi-agent iPLAN framework for the joint design of IWN and new wireless-friendly IE.}
     \vspace{-1mm}
    \label{multi-agent}
\end{figure*}

\textbf{Knowledge:}
% objective information
The knowledge aims to incorporate domain-specific information or expert insights to improve the inference capability of LLM optimizers.
In the context of IWN planning, that may entail relevant technical landmark textbooks, academic papers, historical datasets, wireless channel modeling methods, 3GPP standards, device handbooks, online searching, or expert feedback that provides a contextual theoretical foundation.
By aligning prior multi-modal knowledge with the LLM's decision-making process, a stronger connection is established with the current \textit{status quo} in IWN planning, ensuring the quality and reliability of the proposed solutions for the underlying optimization tasks.

\textbf{Perception:}
Indoor perception aspires to continuously improve the performance of LLM optimizers through interactions with the environment and multi-modal feedback from users, such as diverse KPIs, radio coverage maps, or waveform diagrams.
% external tool 
Expert assessments can be continuously relayed back to LLM optimizers by leveraging external sources of information to evaluate user experience, \eg reported channel state information and the outputs of the IWN simulator.
Hence, LLM-powered optimizers can maintain long-term effectiveness and update their reasoning capacity within a feedback loop, accommodating both static and dynamic environments.

\subsection{Multi-Agent iPLAN Framework for the Joint IWN and New Wireless-Friendly IE Design}\label{Sec: No_Pre_Existing}

The joint IE and IWN design of new wireless-friendly buildings introduces additional degrees of freedom along with conflicting KPIs, rendering it more challenging.
%$, the task always becomes extremely complex with balances to address inside IWN and IE, as well as between them, especially in large-scale buildings. 
Therefore, iPLAN adopts a multi-agent strategy that leverages a step-by-step approach, as illustrated in Fig. \ref{multi-agent}. 
Motivated by \cite{indoor_design}, the joint design task is decomposed into five core sub-tasks. 
Each sub-task is treated as an independent IWN planning problem, characterized by its own set of variables, constraints, and objectives, and is managed by a dedicated iPLAN agent.
% Information Integration
These agents are only required to acquaint themselves with a single sub-task description and exploit domain-specific knowledge relevant to their respective sub-task.
Finally, the generated solutions are integrated and collaboratively evaluated, providing feedback to balance and update the agents' performance.
Leveraging this multi-agent iPLAN framework, agents can focus on addressing specific requirements within each sub-task while ensuring a coordinated multi-agent balance across the whole system.
The detailed design of agents is presented as follows:

\begin{figure*} [t]
    \centering
    \includegraphics[width = 0.85\textwidth]{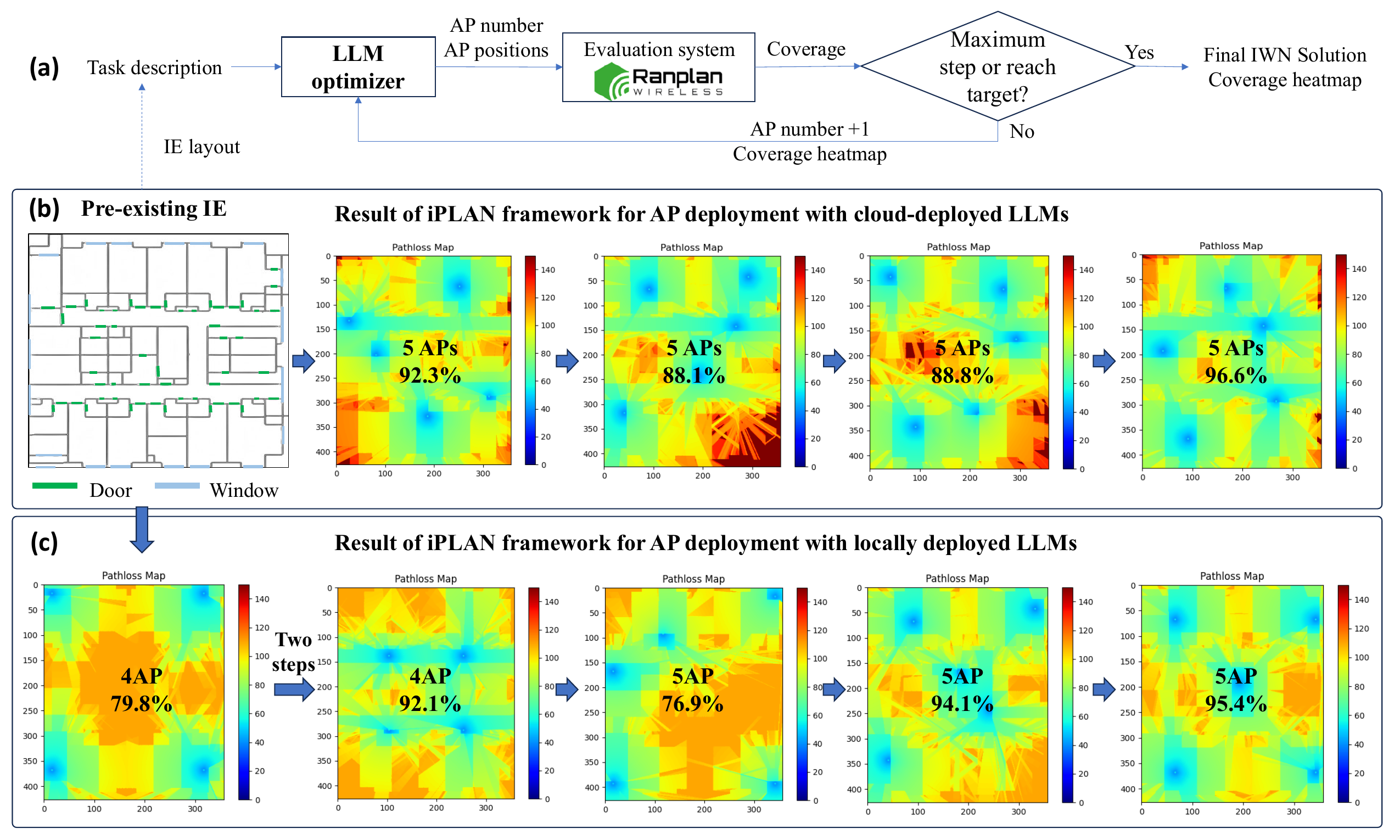}
    \vspace{-2mm}
    \caption{The iPLAN framework and result for AP deployment in pre-existing IE.}
    \label{AP_P-EIE}
     \vspace{-4mm}
\end{figure*}

% high-level 一点？？  合一合

\textbf{1. Layout Agent:} 
A fundamental step in IE design is defining the building contour.
 As illustrated in Fig.~\ref{multi-agent}, we use a layout agent that generates the location, material, and thickness of the external building layout and the interior walls' spatial arrangement based on the IE task description and architectural knowledge.

\textbf{2. Entity Agent:} 
According to the generated IE layout, the entity agent aspires to place key indoor entities, such as windows, doors, furniture, and columns, specifying their locations, height, material, and size.
In addition, by referring to design manuals, historical blueprints, and relevant building regulations, the entity agent ensures that the deployed structural entities comply with architectural standards. 

\textbf{3. IWN Design Agent:}
The IWN design agent is responsible for the network deployment and outputs equipment selection and configuration, such as the number of APs, sensors, or reconfigurable intelligence surfaces (RISs) and their configuration, \eg antenna steering direction or RIS state. That is achieved by conveying essential information as shown in the iPLAN framework.

\textbf{4. Correction Agent:} To amend errors in the optimization process, a dedicated correction agent can be used in order to ensure consistency between the IE and IWN design choices, refining them, if necessary, by adjusting the IWN deployment. For instance, that might entail identifying misplaced wireless equipment and providing relevant modifications. % to ensure optimal integration and performance.

\textbf{5. Evaluation \& Update Agent:} This agent entails two evaluation methods.
% UE 
On one hand, users can evaluate the rationality and aesthetics of the proposed planning based on their own experience and perception.
% tools
On the other hand, external tools are used to evaluate individually whether the IWN and IE goals and constraints are met.
% considering weight
Finally, the agent considers a weighted overall evaluation that depicts the wireless-friendliness of the new building to strike a balance among agents, which is feedback to all agents as a global experience to iteratively update their performance from the global perspective.

\section{Case Study}

To exemplify the potential of LLMs in IWN planning, we consider two use cases for the planning scenarios discussed in the previous section. 
% env
Our experiments were conducted with a 2-Core Intel i5-10505 3.20GHz CPU and 16GB RAM, whilst the LLM agent components leverage ChatGPT 4o, leveraging properly designed prompts\footnote{Available in: https://github.com/Kimboshef/Prompt-for-iPLAN. \label{prompt1}}.

\subsection{AP Deployment in Existing IE}
For the first use case, we aim to deploy new APs to reach a coverage target of 95\% in a pre-existing complex IE with glass windows and wooden doors. We consider the coverage to be satisfactory if the pathloss in this location is smaller than 110 dBm.
As shown in Fig. \ref{AP_P-EIE}(a), for this task, iPLAN receives the IE layout figure, a task description, and essential knowledge to guide the LLM optimizer, using the prompt engineering presented in  our GitHub repository\footnotemark[1], %method \footnote{Available in: https://github.com/Kimboshef/Prompt-for-iPLAN \label{prompt1}} 
to suggest the number and positions of APs \cite{LLM_AP_deploy}. That allows us to adhere to the description, knowledge, and perception modules discussed in Section~\ref{sec:Pre_Existing}. Then, the ``evaluator",  using a high-performance propagation tool, emulates the IWN performance, and the resulting coverage level and heatmap are fed back to the LLM optimizer to assess the optimization process.  If the stop criterion is not reached,  the LLM optimizer iteratively updates the IWN solution until the maximum steps or coverage target is achieved. For our analysis, we consider a commercial cloud-deployed LLM, ChatGPT 4o, referred to as iPLAN-C, and a locally deployed LLM, DeepSeek-R1-70B, referred to as iPLAN-L. It is noted that iPLAN-L is deployed on local servers and accessed via user interfaces.
Fig. \ref{AP_P-EIE}(b) depicts the optimization process of iPLAN-C, with the coverage level gradually reaching 96.6\%, attesting that the cloud-deployed LLM can effectively identify a high-performing IWN topology. 
Fig. \ref{AP_P-EIE}(c) depicts the same process with iPLAN-L, which achieves 95.4\% coverage, within a slightly increased number of iterations, due to its smaller weight scale.

Importantly, when compared to a conventional meta-heuristic and RL algorithms, namely ant colony optimization (ACO) and proximal policy optimization (PPO), respectively, the iPLAN framework can converge substantially faster, \ie in approximately two orders of magnitude fewer iterations, and attain better coverage performance, as shown in Table~\ref{table:iterations}. Remarkably, unlike these methods, iPLAN does not entail environment-specific design and tuning, \eg selecting training policy for PPO and evaporation rate for ACO.

%the RL method involves substantial pre-design work, such as defining the network architecture, training policy, state-action space, and hyperparameter selection, which becomes challenging in complex environments with an uncertain number of APs.}

% that the LLM optimizer can fully consider the IWN task and pre-existing layout to effectively update the coverage performance from 92.3\% to 88.8\% and finally reach the coverage target with 96.6\% by reflecting through feedback.

\begin{table}
\begin{center}
\setlength{\abovecaptionskip}{1pt}
\setlength{\belowcaptionskip}{1pt}
\renewcommand{\arraystretch}{1.2}
\caption{Comparison between iPLAN frameworks, PPO, and ACO}
\label{table:iterations}
\centering
\begin{threeparttable}
\begin{tabular}{ccccc}
\hline
 Algorithm & \textbf{iPLAN-C\tnote{2}} & \textbf{iPLAN-L\tnote{3}} & \textbf{PPO} & \textbf{ACO} \\\hline
Iterations & 4 & 7 & 406  & 516  \\
Coverage performance & 96.6\% & 95.4\% &  95.1\%  & 95.3\% \\\hline 

\end{tabular}
\begin{tablenotes}   
        \footnotesize             
        \item[2] iPLAN-C leverages cloud-deployed LLM, ChatGPT-4o.
        \item[3] iPLAN-L leverages locally deployed LLM, DeepSeek-R1-70B.
\end{tablenotes}
\end{threeparttable}
\end{center}
\end{table}

\begin{figure*}[t]
    \centering
    \includegraphics[width = 0.90\textwidth]{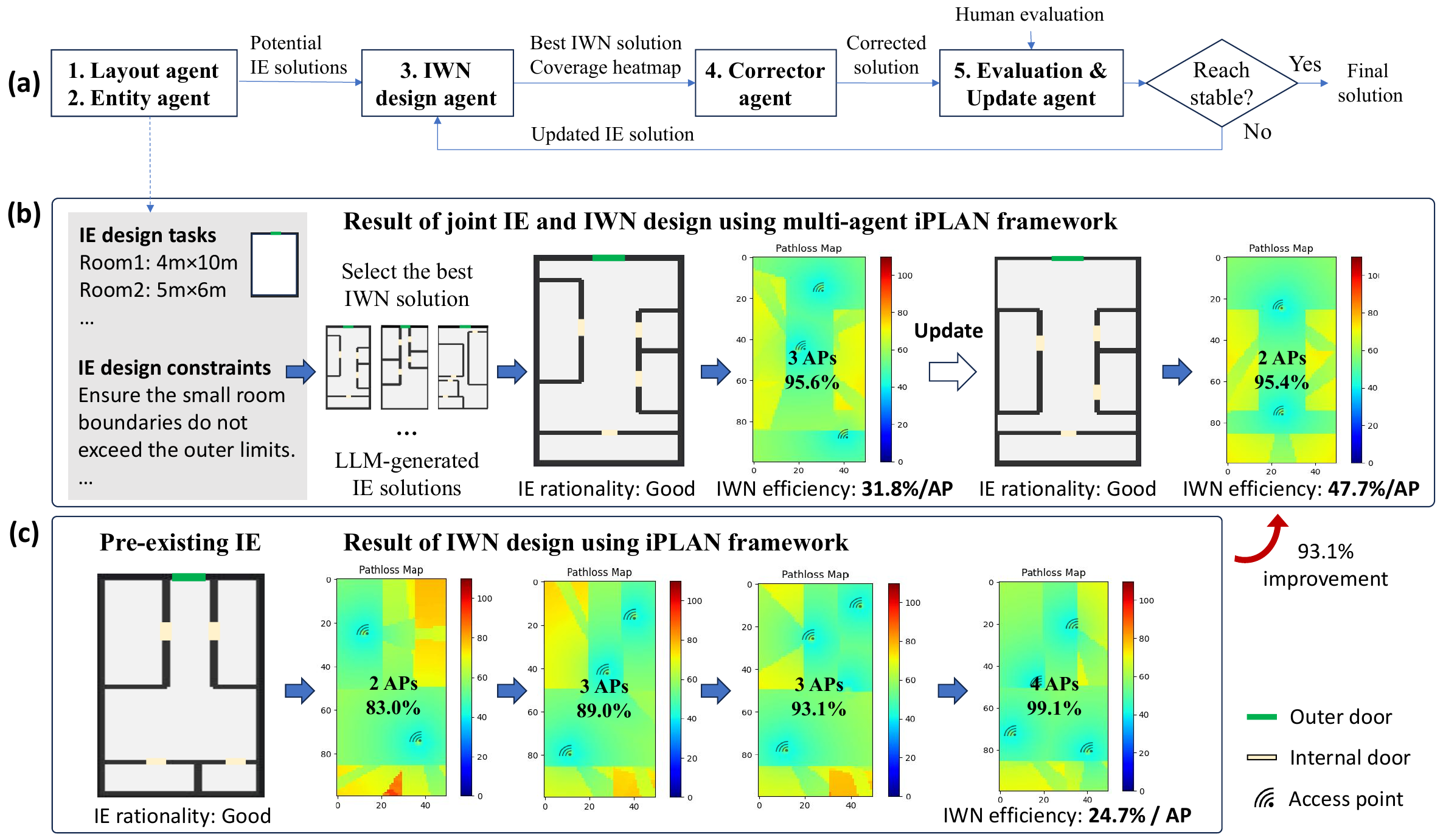}
    \vspace{-2mm}
    \caption{The multi-agent iPLAN framework and results of joint design of building layout and AP deployment, compared with pre-existing IE.}
    \label{AP_result}
\end{figure*}

\subsection{Multi-Agent Joint Design of Wireless-friendly IE and AP Deployment}

For the concurrent IE and IWN design,  we will consider the architectural specifications of a pre-existing building, and we will then task the multi-agent iPLAN to pronounce a more wireless-friendly IE.  To this end, we first identify the building wall and door layout that yields good wireless performance \cite{paper_zhang2022BWP1} while adhering to common architectural standards and rules before deploying new APs. 
% target room
Our target is a rectangular office space with 20m length and 10m width with an outer door at the center of the width wall. We need to design four small rectangular rooms with fixed sizes inside the office space, each with at least one wall overlapping with the outer walls and a single 0.8m wide wooden door. At the same time, the layout design should ensure aesthetic appeal, indoor circulation, and architectural rationality. Then, we seek to identify an IWN topology that achieves 95\% coverage by adjusting the number of APs and their positions within this office space. We consider the coverage is reached if the pathloss is smaller than 80 dBm for high-quality communication services.

To solve this problem, iPLAN leverages information and reasoning from multiple distinct agents, each assigned to carry out a specific task as per Section~\ref{Sec: No_Pre_Existing}. To properly stipulate the behaviour of each agent, we used different suitably designed prompts\footnotemark[1]. Initially, as shown in Fig. \ref{AP_result}(a), the layout and entity agents propose 10 potential layout design solutions by employing \textit{think step-by-step} prompts that guide chain-of-thought reasoning, based on the input information of the IE design task.
Then, the IWN design agent generates the AP positions and its coverage heatmap, followed by adjustments of the corrector agent according to the constraints described above, as shown in Fig. \ref{AP_result}(c). Note that the IWN design agent assumes the structure presented in Fig. \ref{AP_P-EIE}(a). Consequently, the evaluation and update agent updates the IE solution based on its coverage performance and IE rationality until the process converges or the maximum number of iterations is reached.

% result
Fig.\ref{AP_result}(b) presents a more wireless-friendly design of the pre-existing IE, along with the attained coverage and pathloss heatmaps generated via ray-tracing. For the pre-existing IE, we observe that the IWN network agent can gradually improve coverage performance and finally achieve 99.1\% with 4 APs. Defining the IWN efficiency as the coverage level attained divided by the number of APs, we can observe that the multi-agent iPLAN framework can achieve a 93.1\% IWN efficiency improvement by designing a wireless-friendly IE. Indeed, the best-performing IE design solution proposed by LLM can achieve 95.6\% with 3 APs, corresponding to an IWN efficiency equal to 24.5\%. On the other hand, after the evaluation and update phase in iPLAN, the coverage equals 95.4\% with 2 APs, almost doubling the IWN efficiency by iteratively optimizing the position of rooms and doors. 

Consequently, it is evident that building wireless performance can be improved by adjusting the layout, door placement, and room junctions.
% results
This result shows that significant optimization space exists in the wireless performance of IEs, which can serve as an important reference for architects.

\section{Conclusion and Open Challenges}    % and Open Challenges

The emerging concept of LLM-driven optimization instates a new perspective on IWN planning, inaugurating a new class of network planning tools that will underpin next-generation fully automated and intelligent networks. In this article, to give prominence to this transformative paradigm, we outlined key challenges related to IWN planning and showcased how LLM optimizers can be integrated into a holistic optimization framework. We then presented how this general approach can be adapted to solve prevalent planning problems related to IWNs installation in pre-existing buildings, as well as the joint IE and IWN design, ensuring the construction of wireless-friendly buildings. Our vision of LLM-driven IWN planning is founded upon these two fundamental deployment concepts and the numerical results presented that attest to the potential of LLM-driven optimizers to supplant conventional techniques.

% open challenges 
% privacy
Yet, this promising paradigm gives rise to new concerns, and we identify three major challenges on the roadmap ahead. First, IWN planning often involves sensitive data, such as building layouts, user locations, and device usage preferences, raising potential confidentiality risks.
% complex env / 3D environments / dynamic     
Second, the computational complexity and communication overhead of LLMs present obstacles when adapting iPLAN to three-dimensional (3D) or dynamic IEs. Third, it is essential to assess the financial and energy feasibility of wide-scale adoption of LLMs to support such tasks.
% % 3D
% 3D IWN planning requires a more fine-grained representation and additional info such as the height of devices, antenna radiation patterns, and 3D models.
% % dynamic
% user mobility and traffic requirements, \eg dynamic beamforming and power adaptation.
% future works   
To mitigate these challenges, our future works involve locally-hosted proprietary lightweight LLMs for data protection, financial cost-saving, and domain-specific fine-tuning, although this requires additional computational resources and maintenance.
% management enhancement
Furthermore, the integration of iPLAN frameworks with network management systems will be studied to enhance service orchestration \& automation.

\section*{Acknowledgment}

The work of S. Bakirtzis is supported by the Foundation for Education and European Culture. This work is supported in part by the Eureka COMET (with funding from Innovate UK, No. 10099265) and Horizon Europe COVER project, No. 101086228 (with funding from UKRI grant EP/Y028031/1). K. Wang acknowledges the support by the Royal Society Industry Fellowship (IF$\setminus$R2$\setminus$23200104).

% corresponding author

\bibliography{Reference}
\bibliographystyle{IEEEtran}

% \section{Biographies}
% \vspace{-15 mm}
% \printbibliography  
 \vspace{-13 mm}
\begin{IEEEbiography} {Jinbo Hou} is pursuing the Ph.D. degree with the Department of Electronic and Electrical Engineering of the University of Sheffield. Jinbo is also
with Brunel University of London.
\end{IEEEbiography}

\vspace{-12mm}

\begin{IEEEbiography} {Stefanos Bakirtzis}
 (M) received the Ph.D. degree in Computer Science from the University of Cambridge, where he is currently a research associate. He has received the Marie Skłodowska-Curie Actions-Innovative Training Networks Doctoral Fellowship and the Onassis Foundation Scholarship.
 % He is currently working on the Big Data Analytics for Radio Access Networks (BANYAN) project as a member of Ranplan Wireless with the University of Cambridge.

\end{IEEEbiography}
\vspace{-12.5 mm}

   \begin{IEEEbiography} {Kehai Qiu} 
(SM) is working toward the Ph.D. degree with the Computer Laboratory, University of Cambridge, Cambridge, U.K. He received the Marie Skłodowska-Curie Actions-Innovative Training Networks (MSCA-ITN) Fellowship.
 
\end{IEEEbiography}
\vspace{-12.5 mm}

 % He is currently working on the Big Data Analytics for Radio Access Networks (BANYAN) project as a member of Ranplan Wireless with the University of Cambridge.
  \begin{IEEEbiography} {Sichong Liao} is pursuing the Ph.D. degree with the Department of Electronic and Electrical Engineering of the University of Sheffield. 

\end{IEEEbiography}
  \vspace{-12.5 mm}

\begin{IEEEbiography} {Hui Song}
is the Co-Founder and CTO of Ranplan Wireless, Cambridge, U.K. Through the collaboration with academic and industry partners, he is a Principal Investigator or a Co-Investigator (Co-I) of 15 research projects funded by EC FP7/H2020, Eurostars and Innovate U.K. %These projects laid the foundation for Ranplan Professional and DECADE was selected for entry onto EC’s Innovation Radar platform. %As a Co-I of Eurostars Build-Wise project, a ground-breaking building wireless performance evaluation metrics were proposed for the first time. 

\end{IEEEbiography}
  \vspace{-12.5 mm}

\begin{IEEEbiography} {Haonan Hu} (M) is currently a Postdoctoral Fellow in the Department of Electronic and Electrical Engineering, University of Sheffield, UK.
 
\end{IEEEbiography}
 \vspace{-12.5 mm}

\begin{IEEEbiography} {Kezhi Wang} is working as a Professor with the Department of Computer Science, Brunel University of London, U.K. His research interests include wireless communications, mobile edge computing, and machine learning.
 
\end{IEEEbiography}
 \vspace{-12.5 mm}

\begin{IEEEbiography} {Jie Zhang}
(SM) is with the R\&D Department, Cambridge AI+ Ltd., U.K. He is also the Founder, Board Director, and CSO of Ranplan Wireless.
 
\end{IEEEbiography}
\end{document}